\documentclass[aps,prd,superscriptaddress,showpacs, twocolumn]{revtex4-1}
\usepackage{graphicx}
\usepackage{epstopdf}
\usepackage{amsmath}
\usepackage{amsfonts}
\usepackage{amssymb}
\usepackage{latexsym}
\usepackage{hyperref}
\usepackage[english]{babel}
\usepackage[utf8]{inputenc}
\usepackage[colorinlistoftodos]{todonotes}
\usepackage{xcolor}
\usepackage{slashed}
\usepackage{bm}  

\def \and{\textmd{and}}

\begin{document}
\title{On a Five-Dimensional Chern-Simons AdS Supergravity without gravitino}

\author{Y.M.P. Gomes}\email{ymuller@cbpf.br}
\affiliation{Centro Brasileiro de Pesquisas F\'{i}sicas (CBPF), Rua Dr Xavier Sigaud 150, Urca, Rio de Janeiro, Brazil, CEP 22290-180}

\author{J. A. Helayel-Neto}\email{helayel@cbpf.br}
\affiliation{Centro Brasileiro de Pesquisas F\'{i}sicas (CBPF), Rua Dr Xavier Sigaud 150, Urca, Rio de Janeiro, Brazil, CEP 22290-180}

\begin{abstract}
Based on recent discussions on the so-called unconventional supersymmetry, we propose a 5D Chern-Simons AdS-$\mathcal{N}$-SUGRA formulation without gravitino fields and show that a residual local SUSY is preserved. We explore the properties of CS theories to find a solution to the field equations in a 5D manifold. With a Randall-Sundrum-type ansatz, we show that this particular dimensional reduction is compatible with SUSY, and some classes of 4D solutions are then analyzed. 
\end{abstract}

\pacs{04.65.+e,11.15.Yc}
\maketitle


 \section{Introduction} \label{sec_intro}

\indent 
An alternative method to build up a theory with SUSY is by implementing a gauge theory for a super-algebra that includes an internal gauge group, $\mathcal{G}$, along with a local $SO(1,D-1)$ algebra that has to be set up to connect these two symmetries through fermionic supercharges \cite{uncon,uncon1,uncon2}. In these references, the field multiplet is composed by a (non-)Abelian field, A, a spin-1/2 Dirac fermion, $\psi$, the spin connection, $\omega^{ab}$, the d-bien, $e^a$, and additional gauge fields which  complete the degrees of freedom to accomplish the supersymmetrization. These additional fields are dependent on the structure of the group and the space-time which we intend to work in. The representations of the fields are not all the same. The Dirac spinor transforms under the fundamental representation, while the gauge connection belongs to its adjoint representation of $\mathcal{G}$. In this framework, the metric is completely invariant under the symmetries $\mathcal{G}$, $SO(1,D-1)$ and supersymmetry.
\paragraph*{}Due to the properties above, the model displays important differences in comparison with standard SUSYs. For example, there is no superpartners with degenerate masses, nor an equal number of degrees of freedom of bosons and fermions. There is not even a spin-3/2 fermion, i.e., a gravitino, in the spectrum of the model \cite{uncon,uncon1,uncon2}.
\paragraph*{} It is remarkable that, in odd dimensions, the Chern-Simons (CS) form is quasi-invariant under the whole supergroup. On other hand, for even dimensions, the symmetry breaks into $\mathcal{G} \times SO(1,D-1)$. For example, for $D=4$, the super-group can have no invariant traces, and this is the reason why supersymmetry breaks down. The action in four dimensions must be seen as an effective description, due to, for instance, a quartic fermionic coupling that shows up and prevents the model from being renormalizable \cite{uncon,uncon1,uncon2}. 
\paragraph*{}The paradigm that the procedure still keeps from standard SUSY is that fermion and bosons can be combined into a unique non-trivial representation of a supergroup. The differences appear already in the scenario where the SUSY works. In this proposal, SUSY is an extension of the symmetries of the {\it tangent space}. Since Dirac fermions are in the $[(\frac{1}{2},0)\oplus(0,\frac{1}{2})]$-representation of Lorentz group, SUSY is implemented as an extension of the tangent space symmetries. This approach allows us to implement SUSY in any manifold,by looking for the symmetries of the tangent bundle. Another difference is found in the field representations \cite{uncon}.
\paragraph*{}A Chern-Simons $AdS_5$ supergravity is a gauge model based on a SUSY extension of the $AdS_5$ gravity. Based on the no-gravitini approach \cite{uncon} and on the structure of $SO(4,2)$ group, we work with a field that is a 1-form gauge connection \cite{sugcs2}:
\begin{equation}
\hat{A} = \hat{e}^a J_a + \frac{1}{2} \hat{\omega}^{ab}J_{ab} + \hat{A}^k T_k + (\bar{\psi}^r\hat{\Gamma}Q_r + \bar{Q}^r \hat{\Gamma} \psi_r) + \hat{b} \mathbb{K},
\end{equation}
\paragraph*{}where the hat stands for 5-dimensional forms; $\hat{\Gamma} = \hat{e}^a \gamma_a$, with $a,0,...,4$; $k=1,...,\mathcal{N}^2-1$ and $r=1,...,\mathcal{N}$. This 1-form has values in the $SU(2,2|\mathcal{N})$ super-algebra, whose bosonic sector is given by $SU(2,2) \otimes SU(\mathcal{N}) \otimes U(1)$, where $SU(2,2) \simeq SO(4,2)$ \cite{sugcs2}.

\paragraph*{}The infinitesimal gauge transformation is given by $\delta \hat{A} = \hat{d} \epsilon + [ \hat{A}, \epsilon\}$, with $\epsilon = \epsilon^a J_a + \frac{1}{2} \epsilon^{ab} J_{ab} + \epsilon^k T_k + \bar{\chi}^rQ_r + \bar{Q}^r \chi_r + \epsilon_b \mathbb{K}$. In components, we have:
\begin{subequations}
\begin{equation}
\delta \hat{e}^a = \hat{d}\epsilon^a + \hat{\omega}^{ab} \epsilon_b + \epsilon^{ab} \hat{e}_b+\frac{1}{2}(\bar{\psi}^r \hat{\Gamma} \gamma^a \chi_r +  \bar{\chi}^r\gamma^a \hat{\Gamma} \psi_r) ~,
\end{equation}
\begin{equation}\delta \hat{\omega}^{ab} = \hat{d}\epsilon^{ab} + \hat{\omega}^{ac} \epsilon_c^{~b} +\hat{\omega}^{bc} \epsilon_c^{~a} + \frac{1}{4}(\bar{\psi}^r \hat{\Gamma} \gamma^{ab} \chi_r +  \bar{\chi}^r\gamma^{ab} \hat{\Gamma} \psi_r)~,
\end{equation}
\begin{equation} \delta \hat{A}^k = \hat{d} \epsilon^k + f^{k}_{~lm}\hat{A}_l \epsilon^m - i (\bar{\psi}^r (\tau^k)_r^{~s}\hat{\Gamma} \chi_s +  \bar{\chi}^r \hat{\Gamma} (\tau^k)_r^{~s}\psi_s)~,
\end{equation}
\begin{equation}\delta( \hat{\Gamma} \psi_r) = \hat{\vec{\nabla}} \chi_r\end{equation}
\begin{equation}\delta \hat{b} = \hat{d}\epsilon_b + i (\bar{\psi}^r \hat{\Gamma} \chi_r +  \bar{\chi}^r\hat{\Gamma} \psi_r)~,
\end{equation}
\end{subequations}
\paragraph*{}where  $\hat{\vec{\nabla}} \chi_r = \hat{d}\chi_r +[i(\frac{1}{4}- \frac{1}{\mathcal{N}})\hat{b} + \frac{1}{2}  \hat{e}_a \gamma^a+ \frac{1}{4} \hat{\omega}_{ab} \gamma^{ab}]\chi_r + \hat{A}_k (\tau^k)_r^{~s}\chi_s$. The field-strength is given by $\hat{F} = \hat{d} \hat{A} + \frac{1}{2}[\hat{A}, \hat{A}\}$. In components, we have $\hat{F} = \hat{F}^a J_a + \frac{1}{2}\hat{F}^{ab}J_{ab} + \hat{F}^k T_k + \bar{\hat{\Theta}}^r Q_r + \bar{Q}^r \hat{\Theta}_r + F \mathbb{K}$, where:
\begin{subequations}

\begin{equation}
\hat{F}^{a} = \hat{d} \hat{e}^a +  \hat{\omega}^{a}_{~b} \hat{e}^b  + \bar{\psi}^r\hat{\Gamma}  \gamma^a \hat{\Gamma} \psi_r = \hat{D}_{\hat{\omega}} \hat{e}^a + \bar{\psi}^r\hat{\Gamma}  \gamma^a \hat{\Gamma} \psi_r~, 
\end{equation}
\begin{equation}\hat{F}^{ab} = \hat{R}^{ab} + \hat{e}^a \hat{e}^b + \frac{1}{2}\bar{\psi}^r \hat{\Gamma}  \gamma^{ab} \hat{\Gamma} \psi_r ~,
\end{equation}
\begin{equation}
\hat{F}^k = \hat{d} \hat{A}^k +  f^{k}_{~lm}\hat{A}^{l} \hat{A}^m  + \bar{\psi}^r\hat{\Gamma} 
(\tau^k)_r^{~s}\hat{\Gamma} \psi_s~,
\end{equation}
\begin{equation}
\hat{\Theta}_r = (\hat{\vec{\nabla}})_r^s(\hat{\Gamma}\psi_s) \text{~~,~~} \bar{\hat{\Theta}}^r =  (\hat{\vec{\nabla}})_r^s(\bar{\psi}^s\hat{\Gamma})~,
\end{equation}
\begin{equation}
\hat{F} =  \hat{d} \hat{b} + i \bar{\psi}^r \hat{\Gamma} \hat{\Gamma} \psi_r ,
\end{equation}
\end{subequations}
\paragraph*{}where $\hat{R}^{ab} = \hat{d}\hat{\omega}^{ab}+\hat{\omega}^{ac}\hat{\omega}_c^{~b}$. In the sequel, we shall specifically analyze the SUSY transformations and see how the gravitino sector is suppressed from the model.
\section{SUSY transformation}
\paragraph*{}In the work of Ref.\cite{uncon1}, to ensure that no gravitini appear in the spectrum in 3D action, the authors show that the dreibein remains invariant under gauge and supersymmetry transformations, but rotates as a vector under the Lorentz subgroup. To do this, we must look for the SUSY transformations. In the fermionic part, we have $\delta( \hat{\Gamma} \psi_r) = \hat{\vec{\nabla}} \chi_r$, where $\chi$ is the local SUSY parameter. Any vector with spinor index can be split into irreducible representations: $1 \otimes 1/2 = 3/2 \oplus 1/2$ of the Lorentz group. So, for $\xi_a^\alpha =( P_{3/2} + P_{1/2} )_b^a\xi_b^\alpha = \phi_a^\alpha + \Psi_a^\alpha$, where $(P_{3/2})_a^b = \delta_a^b - \frac{1}{5}\gamma_a \gamma^b = \delta_a^b - (P_{1/2})_a^b$ are the projectors, $\phi_a^\alpha$ are the 3/2-component and $\Psi_a^\alpha$ is the 1/2-component. Therefore, we have $(P_{3/2})_a^b \gamma_b \psi = 0$, by definition. SUSY transformation yields:
\begin{equation}\label{etransf}
\delta( \hat{\Gamma} \psi_r) = \delta \hat{e}^a \gamma_a \psi_r + \hat{e}^a \gamma_a \delta \psi_r = \hat{\vec{\nabla}} \chi_r.
\end{equation}
\paragraph*{}Applying the $P_{3/2}$-projector to the equation above, we find that \begin{equation}(P_{3/2})_\mu^{~\nu} \hat{\nabla}_\nu \chi_r =0~,\end{equation} which implies that $\hat{\nabla} \chi_r = \hat{e}^a \gamma_a \rho_r$, for an arbitrary spinor $\rho$. This condition guarantees that the symmetry transformations close off-shell without the need of introducing auxiliary fields \cite{uncon1}. Applying $P_{1/2}$-projector to the equation \eqref{etransf} we obtain that, under SUSY, $\delta \psi_r = \rho_r$ and $\delta \hat{e}^a = 0$. The spinor $\rho_r$ obeys the Killing equation; the number of Killing spinors defines the number of unbroken supersymmetries, i.e., supersymmetries respected by the background \cite{uncon1}. For instance, if $\rho_r = 0$, we have $\chi_r = \text{constant}$ (covariantly constant), and we obtain a global SUSY. For a general solution, a Hamiltonian analysis must be carried out to extract the exact solution for the SUSY parameter \cite{uncon2}. 
\section{5D topological Action }\label{sec3}
\paragraph*{}The topological action can be written as a Chern-Simons action in 5 dimensions \cite{sugcs2}:
\begin{equation}
S^{5D} = \int \langle \mathcal{A}  \mathcal{F}  \mathcal{F} - \frac{1}{2} \mathcal{F}\mathcal{A}\mathcal{A}\mathcal{A} + \frac{1}{10}\mathcal{A}\mathcal{A}\mathcal{A}\mathcal{A}\mathcal{A} \rangle ,
\end{equation}
\paragraph*{}where $\langle ... \rangle$ stands for the supertrace. The only non-vanishing supertraces are:
\begin{equation}\nonumber
\langle J_a J_{b c} J_{d e} \rangle = -\frac{1}{2} \varepsilon_{abcde} ~~,~~  \langle T^i T^j T^k \rangle  = - f^{ijk}
\end{equation}
\begin{equation}\nonumber
\langle \mathbb{K} J_{a b} J_{c d} \rangle = -\frac{1}{4}\eta_{ab,cd} ~~,~~  \langle \mathbb{K} T^i T^j \rangle  = - \frac{1}{\mathcal{N}} \delta^{ij}
\end{equation}
\begin{equation}\nonumber
\langle \mathbb{K} J_{a } J_{b} \rangle = -\frac{1}{4}\eta_{a b} ~~,~~  \langle \mathbb{K}\mathbb{K}\mathbb{K} \rangle = \frac{1}{\mathcal{N}^2} + \frac{1}{4^2} 
\end{equation}
\begin{equation}\nonumber
\langle \bar{Q}^\alpha_r J_{a b} Q_\beta^s \rangle = -\frac{i}{4}(\Gamma_{ab})^\alpha_\beta  \delta^s_r ~~,~~  \langle \bar{Q}^\alpha_r T^i Q_\beta^s \rangle  = -\frac{i}{2}\delta^{\alpha}_\beta (\tau^i)^s_r
\end{equation}
\begin{equation}\nonumber
\langle \bar{Q}^\alpha_r J_{a} Q_\beta^s \rangle = -\frac{i}{2}(\Gamma_{a})^\alpha_\beta  \delta^s_r ~~,~~  \langle \bar{Q}^\alpha_r \mathbb{K} Q_\beta^s \rangle  = -\frac{1}{2}(\frac{1}{4} + \frac{1}{\mathcal{N}})\delta^{\alpha}_\beta \delta^s_r
\end{equation}
\paragraph*{}Using these definitions, we can find the components of the action $S^{5D} = S_G + S_{SU(\mathcal{N})} + S_{U(1)} + S_f$, where:
\begin{equation}
S_G = -\frac{1}{2}\epsilon_{abcde} \int \hat{F}^{ab} \hat{F}^{cd} \hat{e}^e - \frac{1}{2} \hat{F}^{ab} \hat{e}^c\hat{e}^d\hat{e}^e + \frac{1}{10} \hat{e}^a\hat{e}^b\hat{e}^c\hat{e}^d\hat{e}^e~,
\end{equation}

\begin{eqnarray}\nonumber
S_{SU(\mathcal{N})} &=& - \int Tr\Big{[}{\bf \hat{A}} {\bf \hat{F}}  {\bf \hat{F}} - \frac{1}{2} {\bf \hat{A}}{\bf \hat{A}} {\bf \hat{A}} {\bf \hat{F}} + \frac{1}{10}{\bf \hat{A}}{\bf \hat{A}}{\bf \hat{A}}{\bf \hat{A}}{\bf \hat{A}}\Big{]} +\\&& +\frac{i}{2} {\bf \hat{A}} \cdot \bar{\psi}^r\hat{\Gamma} (\hat{\nabla} {\bf \tau} \hat{\nabla})_r^{~s} \hat{\Gamma} \psi_s ~,
\end{eqnarray}

\begin{eqnarray}\nonumber
S_{U(1)} &=& \int (\frac{1}{\mathcal{N}^2} + \frac{1}{4^2}) \hat{b} (\hat{F})^2 + \hat{b}\Big{(}-\frac{1}{4}\hat{F}^{ab}  \hat{F}_{ab} - \frac{1}{4} \hat{F}^a\hat{F}_a+\\\nonumber
&& - \frac{1}{\mathcal{N}}\hat{F}^i \hat{F}_i + \frac{1}{2}(\frac{1}{4}+ \frac{1}{\mathcal{N}})\bar{\psi}^r \hat{\Gamma} (\hat{\nabla}^2)_r^{~s} \hat{\Gamma} \psi_s\Big{)}~,\\
\end{eqnarray}

\begin{equation}
S_f = i \int \bar{\psi}^r \hat{\Gamma}\hat{\mathcal{R}}_{ r}^{~s}(\hat{\nabla} \hat{\Gamma} \psi)_s  + c.c.~,
\end{equation}
\paragraph*{}Here, \begin{eqnarray}\nonumber
(\hat{\nabla}^2)_r^s &=& \big{[}\frac{1}{4}(\hat{R}^{ab} + \hat{e}^a\hat{e}^b)\gamma_{ab} + \frac{1}{2}\hat{T}^a \gamma_a + i (\frac{1}{4}-\frac{1}{\mathcal{N}})\hat{d}\hat{b}\big{]}\delta_{r}^s + \\&&+\big{[}\hat{d}\hat{A}^k  + f^{kk'k''}\hat{A}^{k'} \hat{A}^{k''}\big{]}(\tau^k)_r^s~,
\end{eqnarray}
\begin{equation}
(\hat{\nabla} \tau^k \hat{\nabla})_r^s = (\hat{\nabla}^2)_r^{s'}(\tau^k)_{s'}^s ,
\end{equation}
 and \begin{equation}\hat{\mathcal{R}}_r^{~s} = \big{[}- \frac{1}{4} \hat{F}^{ab} \gamma_{ab} - \frac{1}{2} \hat{F}^a \gamma_a + \frac{i}{2}(\frac{1}{4}+ \frac{1}{\mathcal{N}})\hat{F}\big{]}\delta_r^{~s} + \hat{F}^i (\tau_i)_r^{~s}~.\end{equation}
 \paragraph*{}It should be noticed that, since $\hat{\mathcal{R}}_r^s \supset \hat{\Gamma} \hat{\Gamma}\delta_r^s$, the fermionic part of $S_f$ generates a Dirac-like action for the fermions ($S_f \supset \int d^5 x \bar{\psi}^r \slashed{D} \psi_r$). Notice that the bosonic part is almost the same in comparison with the usual 5D AdS-SUGRA action \cite{sugcs2}. The important difference lies in the fermionic sector.
\subsection{Gauge transformation and the field equations} 
\paragraph*{}The CS 5D action transforms under a gauge transformation as $\delta S^{5D} = \int \langle \mathcal{F} \mathcal{F} \delta \mathcal{A} \rangle$. We can see that due to this identity, one can readily find the field equations in terms of component fields, and they are given by: 
\begin{equation}
\delta \hat{e}^a \rightarrow -\frac{1}{2}\varepsilon_{a b c d e} \hat{F}^{b c} \hat{F}^{d e} -\frac{1}{4} \hat{F}_{b} \hat{F} - \frac{i}{2} \bar{\hat{\Theta}}^r \gamma_a \hat{\Theta}_r =0~,
\end{equation}

\begin{equation}
\delta \hat{\omega}^{ab} \rightarrow -\frac{1}{2}\varepsilon_{a b c d e} \hat{F}^{c d} \hat{F}^e -\frac{1}{4} \hat{F}_{ab} \hat{F} - \frac{i}{2} \bar{\hat{\Theta}}^r \gamma_{ab} \hat{\Theta}_r =0~,
\end{equation}

\begin{eqnarray}\nonumber
\delta \hat{b} &\rightarrow& -\frac{1}{4} \hat{F}^{ab} \hat{F}_{ab} -\frac{1}{4} \hat{F}^{a} \hat{F}_a -\frac{1}{\mathcal{N}}\hat{F}^i \hat{F}_i +(\frac{1}{\mathcal{N}^2}-\frac{1}{4^2}
)(\hat{F})^2+\\&& -\frac{1}{2}(\frac{1}{4}- \frac{1}{\mathcal{N}}) \bar{\hat{\Theta}}^r \hat{\Theta}_r=0~,
\end{eqnarray}

\begin{equation}
\delta \hat{A}^{i} \rightarrow  f^{ikj}\hat{F}^{j} \hat{F}^k +\frac{1}{\mathcal{N}} \hat{F}_i \hat{F} + \frac{i}{2} \bar{\hat{\Theta}}^r (\tau^i)_r^s \hat{\Theta}_s = 0~,
\end{equation}

\begin{equation}
\mathcal{R}_r^{~s} \hat{\Theta}_s  = 0~.
\end{equation}
\paragraph*{}It can be checked that $\mathcal{F}= 0$ is a solution to the field equation. Let us analyze this solution. In components, we have:
\begin{subequations}

\begin{equation}
\hat{F}^{a} = 0 \rightarrow  \hat{T}^a=\hat{D}_{\hat{\omega}} \hat{e}^a =- \bar{\psi}^r\hat{\Gamma}  \gamma^a \hat{\Gamma} \psi_r 
\end{equation}
\begin{equation}
\hat{F}^{ab} = 0 \rightarrow \hat{R}^{ab} + \hat{e}^a \hat{e}^b = - \frac{1}{2}\bar{\psi}^r \hat{\Gamma}  \gamma^{ab} \hat{\Gamma} \psi_r 
\end{equation}
\begin{equation}
\hat{F}^k = 0 \rightarrow \hat{d} \hat{A}^k +  f^{k}_{~lm}\hat{A}^{l} \hat{A}^m  =- \bar{\psi}^r\hat{\Gamma} 
(\tau^k)_r^{~s}\hat{\Gamma} \psi_s
\end{equation}

\begin{equation}\hat{F} = 0 \rightarrow \hat{d} \hat{b} =- i \bar{\psi}^r \hat{\Gamma} \hat{\Gamma} \psi_r.
\end{equation}
\end{subequations}

\paragraph*{}It is useful to pay attention to some special structures. For instance, if we take the 3-form $\hat{S} = \hat{e}_a \hat{T}^a$, we have that $\hat{S} = -\bar{\psi}^r \hat{\Gamma} \hat{\Gamma} \hat{\Gamma} \psi_r = i\star \hat{d} \hat{b}$, where $\star$ represents the Hodge dual in the 5D Manifold. However, by using the Cartan identities, we have $\hat{d}(\hat{e}_a \hat{T}^a) = \hat{T}^a \hat{T}_a - \hat{e}_a \hat{e}_b \hat{R}^{ab}$. By virtue of this identity, we find the equation that follows:
\begin{equation}\label{bdin}
i \hat{d} \star \hat{d} \hat{b} =  \hat{T}^a \hat{T}_a - \hat{e}_a \hat{e}_b \hat{R}^{ab}.
\end{equation}

\paragraph*{}On the other hand, by defining the co-derivative $\hat{d}^{\dagger} = \star d \star$, we may introduce a Laplacian operator, $\hat{\Box} = \hat{d}^{\dagger} \hat{d} + \hat{d} \hat{d}^{\dagger}$, and, using a gauge condition $\hat{d}^{\dagger} \hat{b} = 0$, we have 
$\Box \hat{b} =  \star(\hat{T}^a \hat{T}_a - \hat{e}_a \hat{e}_b \hat{R}^{ab}) 
$. Therefore, the $\hat{b}$-field has a dynamics which respects the equation above. In this sense, we can interpret the topological sector as the source of the $U(1)$ field, $\hat{b}$. Going further, if the fünfbein is invertible, we can define the following operation on some n-form, $(\hat{E}_a\rfloor \hat{V}^a) = \hat{E}^{\mu}_a \hat{V}^a_{\mu \mu_1 \mu_2...\mu_{n-1}}dx^{\mu_2}...dx^{\mu_{n-1}}  = \hat{V}$, where $\hat{E}^\mu_a$
is the inverse of the fünfbein, i.e., $(\hat{E}_a\rfloor \hat{e}^b) = \delta^b_a$.
This operation can normally be extended to forms with any number of Lorentz indices. So, we can define the 1-form $\hat{T} = (\hat{E}_a\rfloor \hat{T}^a) = 10 \star (\hat{e}_a \hat{e}_b \hat{R}^{ab})$. Therefore, the equation \eqref{bdin} can be rewritten as
\begin{equation}
\hat{\Box} \hat{b} =  \star(\hat{T}^a \hat{T}_a) - \frac{1}{10} \hat{T}.
\end{equation}
\paragraph*{}We here omit the fermionic structure of the torsion for simplicity. This shows us that the torsion is the only source for the $\hat{b}$-field and it is a propagating excitation in five-dimensional space-time.

\section{Dimensional Reduction}

\paragraph*{}We have that the index $a= 0,...,4= I,4$; where the index $I$ refers to the $SO(1,3)$ Minkowski group. So, the fields can be split into two pieces \cite{chams,neves}.
\begin{equation}
\hat{\omega}^{ab} = \{ \hat{\omega}^{IJ}, \lambda \hat{b}^I\} ~~,~~ \hat{e}^a = \{\hat{e}^I , \hat{e}^4\}~.
 \end{equation}
\paragraph*{}Besides that, we are also interested in considering the action in a 4-dimensional version; so, we must split the coordinates as $x^\alpha = (x^\mu, \chi)$ and the 1-forms can be written as follows below:
\begin{equation}\label{dimred1}
\hat{\omega}^{IJ} = \omega^{IJ} + \omega^{IJ}_\chi d\chi ~~;~~  \hat{b}^I = {b}^I + {b}^I_\chi d\chi  
\end{equation}
\begin{equation}\label{dimred2}
\hat{e}^I = e^I + e^I_\chi d\chi ~~;~~ \hat{e}^4 = e^4 + e^4_\chi d\chi 
\end{equation}
\begin{equation}\label{dimred3}
\hat{b} = b + b_\chi d \chi ~~; ~~ \hat{A}^k = A^k + A^k_\chi d \chi ~.
\end{equation}
\paragraph*{} Since the 5 D gamma-matrices can be split as $\gamma^a = (\gamma^I, \gamma_5)$, we then have:
\begin{equation}
\hat{\Gamma} = \gamma^I e_I + \gamma_5 e^4 + \Big{[}\gamma^I (e_I)\chi + \gamma_5 e^4_\chi \Big{]} d \chi ~.
\end{equation}

\paragraph*{}As we can see, the equation $\mathcal{F}=0$ satisfies the field equations for the topological action. Therefore, we can analyze this solution in terms of the reduced components (see \ref{app}). 
\paragraph*{}{\bf NOTE - Chamseddine Gauge-Fixing:} In the dimensional reduction of the 5D Chamsedinne action to a 4D action for gravity \cite{chams}, it can be shown that we can fix $e^4 = e^I_\chi =b^I = \omega_\chi^{IJ}=0$, due to the condition $\partial_\chi f = 0$, for any field $f$. However, in our case, this is not possible anymore, due to the supersymmetric character of the transformation. If we wish to preserve SUSY, we should not use the Chamsedinne gauge-fixing. The central question is: which is the gauge-fixing that maintains SUSY and switches off the spurious degrees of freedom? Using the Killing equation, we check that $\delta_{SUSY} \hat{e}^a = 0$. Therefore, we can fix, in principle, $e^4 = e^I_\chi = 0$,  but we still cannot fix $b_I$ and $\omega_\chi^{IJ}$.
\paragraph*{}Since the fünfbein does not transform under SUSY, the analysis of the residual transformation of $\hat{e}^a$ gives us some clues about the ansatz we may assume for the fünfbein. One of the possible ansätze is shown in the next Section.   

\section{Randall-Sundrum Dimensional Reduction}\label{RSR}

\paragraph*{}A Randall-Sundrum-like ansatz is proposed with the assumption that the geometry of 5D space-time has the following structure \cite{rand0,rand,rand1}:
\begin{equation}\label{metric4d}
ds^2_{5D} = e^{-2 \sigma(\chi)} g_{\mu \nu}(x) dx^\mu dx^\nu + G(\chi)^2 d\chi^2~. 
\end{equation}
We can translate \eqref{metric4d} in terms of the following fünfbein:
\begin{equation}\hat{e}^{a}_\alpha=\begin{bmatrix}
h^{I}_{~\mu}(x) e^{-\sigma(\chi)} & 0\\
0 & G(\chi)\\
\end{bmatrix},
\end{equation}
\paragraph*{}where $\sigma$ is called conformal function. A special choice and application of this ansatz in standard AdS SUGRA can be viewed in the paper by Garavuso and Toppan \cite{CBPF1}. The 4D metric can be written as $g_{\mu \nu}(x) = \eta_{IJ} h^{I}_{~\mu} h^{J}_{~\nu}$. This ansatz fixes $e^{4} = e^I_\chi = 0$. This can, in principle, be a problem. But, due to the no-gravitini condition, this choice is viable. Going further, the inverse of the fünfbein is given by: 
\begin{equation}\hat{E}_{a}^\alpha=\begin{bmatrix}
(h^{-1}(x) )_{I}^{~\mu}e^{\sigma(\chi)} & 0\\
0 & \frac{1}{G(\chi)}\\
\end{bmatrix},
\end{equation}
\paragraph*{}where we assume that $h^I_{~\mu}$ has an inverse, i.e., $h^I_{~\mu} (h^{-1})^{J\mu} = \eta^{IJ}$. This opens up the opportunity to define the inverse of the vielbein (in 4D). We have that $e^I = e^{-\sigma}h^I_\mu dx^\mu = e^I_\mu dx^\mu$, which implies $e^I_\mu e_{\nu I} = g_{\mu \nu}$. Therefore, we may define $E^\mu_I = e^{\sigma}h^\mu_{~I}$ so that $e^I_\mu E^{\mu}_J = \delta^{I}_{J}$. We can define a similar operation ``$\rfloor$" in 4D, and we can rewrite the identity above as $E^{\mu}_J e^I_\mu  = (E_J \rfloor e^I) = \delta^{I}_{J}$. We shall use this operation from now on, and the 4D character is implicit in the forms without $`` ~\hat{~}~ "$. Now, we can look at the field equations $\mathcal{F} = 0$. This anzatz gives us that the torsion part yields the following equations: 
 \begin{subequations}
 \begin{equation}
 d e^I + \omega^{I}_{~J}e^J  = T^I =  - \bar{\psi}^r \Gamma\gamma^I\Gamma\psi_r~,
 \end{equation}
 \begin{equation} 
-\sigma' e^I + \omega_\chi^{IJ} e_J = - G \bar{\psi}^r \Gamma\gamma^I \gamma_5  \psi_r~, 
\end{equation}
\begin{equation}
\lambda b^I e_I =- \bar{\psi}^r \Gamma\gamma_5 \gamma^I \psi_r e_I~,
\end{equation}
\begin{equation} 
\lambda b_\chi^I e_I= - G \bar{\psi}^r \gamma^I \psi_r e_I ~,
\end{equation}
\end{subequations}
\paragraph*{}where $\sigma' = \partial_\chi \sigma$, and we have used $\gamma_5^2 = 1$. Note that the first equation gives us that the 4D torsion is algebraically solved interms of a fermionic bilinear. In general, the torsion tensor can be written as:
\begin{equation}
T^I_{~JK} = \frac{1}{3}(\delta^I_J t_K - \delta^I_K t_J) + \frac{1}{6}\epsilon^{I}_{~JKL} s^L + q^{I}_{~JK}~,
\end{equation}
\paragraph*{}where the q-tensor are in general discarded. Therefore, from the first identity of the previously set of equations, we see that we may write a 3-form $S$ such that $S = e_I T^I = - \bar{\psi}^r \Gamma\Gamma\Gamma \psi_r$ and a 1-form $T = (e_I \rfloor T^I) = - 4 \bar{\psi}^r \Gamma \psi_r$, both components of the 2-form torsion,  $T^I_{~JK}=(e_K \rfloor e_J \rfloor T^I) = - \bar{\psi}^r\gamma_J\gamma^I\gamma_K\psi_r$, i.e.:
\begin{equation}
 t_I = -\bar{\psi}^r\gamma_I\psi_r~~;~~ s_I = -\bar{\psi}^r\gamma_I\gamma_5 \psi_r ~~;~~q^{I}_{~JK}=0~.
\end{equation}
\paragraph*{}Following this line of arguments, from the second equation, we have that $\sigma'(\chi)  = 4 G(\chi) \bar{\psi}^r \gamma_5 \psi_r$. In other words, we have a direct relation between the conformal function, $\sigma$, the $\chi$-component of the fünfbein, $G(\chi)$, and the chiral scalar bilinear, $ \bar{\psi}^r \gamma_5 \psi_r$. In the case $ \bar{\psi}^r \gamma_5 \psi_r = 0$, the conformal function will be some arbitrary constant. We can directly see that $\omega_\chi^{IJ} = - G \bar{\psi}^r\gamma^{IJ} \gamma_5\psi_r$, $\lambda b^I =- \bar{\psi}^r \Gamma \gamma_5 \gamma^I \psi_r$ and $\lambda b_\chi^I= - G \bar{\psi}^r \gamma^I \psi_r= G t^I$ . The only component of the spin connection that does not come out as a fermionic bilinear is the 4-dimensional spin connection, $\omega^{IJ}$. 
\paragraph*{}Using the equations of motion and the global conformal symmetry, $\hat{e}^a \rightarrow \ell \hat{e}^a$ , $\psi_r \rightarrow \ell^{-1} \psi_r$, present in the connection, we can express the Ricci scalar by means of the following equation:
\begin{eqnarray}\nonumber\label{mainr}
R(\tilde{\omega}) &=&  - \frac{8}{\ell^2} -  (1+ \lambda^2)\ell^2(\bar{\psi}^r \gamma^{IJ} \gamma_5\psi_r)(\bar{\psi}^s \gamma_{IJ} \gamma_5\psi_s) +\\\nonumber
&&- 10 ~ \bar{\psi}^r\psi_r  +\frac{4}{3} \ell^2(\bar{\psi}^r\gamma_I\psi_r)(\bar{\psi}^s\gamma^I\psi_s)  +\\\nonumber
&& + \frac{\ell^2}{24} (\bar{\psi}^r\gamma_I\gamma_5 \psi_r) (\bar{\psi}^s\gamma^I\gamma_5 \psi_s) - 4 \ell^2 (\bar{\psi}^r \gamma_5\psi_r)^2~, \\
\end{eqnarray}
\paragraph*{}where the spin connection is written as $\omega^{IJ} = \tilde{\omega}^{IJ} + K^{IJ}$, with $K^{IJ}$ being the 1-form the contortion, related with the torsion by the relation $T^I = D(\tilde{\omega})e^I + K^I_{~J} e^J = K^I_{~J} e^J$ and $K_I = (E^J \rfloor K_{IJ}) = t_I$. Eq.\eqref{mainr} is the main result of our paper. Independently from the possible dependence of the $\chi$-coordinate on the fermionic field, we may state that our result can be interpreted as an effective cosmological constant where the fermionic matter is distributed. This can also can affect the internal structure of stars, specially the most dense ones. 
\paragraph*{}The torsion components, with the correct dimension, are written as follows:
\begin{equation}
 t_I = -\frac{1}{\ell}\bar{\psi}^r\gamma_I\psi_r~~;~~ s_I = -\frac{1}{\ell}\bar{\psi}^r\gamma_I\gamma_5 \psi_r ~~;~~q^{I}_{~JK}=0
\end{equation}
\paragraph*{}As we have sees in Section \ref{sec3}, the gauge field $\hat{b}$ acquires dynamics in the 5-dimensional space. Now, after dimensional reduction, we can look its components, $\hat{b} = (b, b_\chi= \Phi)$, and we have that:
\begin{eqnarray}\nonumber
\Box b &=& (\frac{3\sigma'-1}{2\ell})(\bar{\psi}^r \Gamma \psi_r)+\\&&+2 G \epsilon_\mu^{~~ \nu \rho \lambda}(\bar{\psi}^r \gamma_\nu \gamma_5 \gamma_\rho \psi_r)(\bar{\psi}^r \gamma_\lambda \psi_r) dx^\mu ~,
\end{eqnarray}

\begin{eqnarray}\nonumber
\Box \Phi &=& - \frac{1}{2 \ell}\bar{\psi}^r\gamma_5 \psi_r  -\epsilon^{\mu \nu \rho \lambda}[(\bar{\psi}^r \gamma_\mu \gamma^I \gamma_\nu \psi_r)(\bar{\psi}^r \gamma_\rho \gamma_I \gamma_\lambda \psi_r)+\\&&+(\bar{\psi}^r \gamma_\mu \gamma_5 \gamma_\nu \psi_r)(\bar{\psi}^r \gamma_\rho \gamma_5 \gamma_\lambda \psi_r)]~,
\end{eqnarray}
\paragraph*{}where $\Box$ here means $\hat{\Box}$ in the framework of the RS dimensional reduction. As one can notice, the vector current, $j^\mu = \bar{\psi}^r \gamma^\mu \psi_r$, is a source to the vector gauge field, $b$, with effective charge $q= \frac{3 \sigma'-1}{2 \ell}$. The pseudo-scalar bilinear act as a source to the component $\Phi$. Besides that, a unusual source, quartic in the fermionic fields, also appears in the equation of motion of both components.  

 \section{Conclusions}

\paragraph*{}We have here presented a 5D Chern-Simons AdS-super-gravity model without a gravitino. The formulation proposed in \citep{uncon} generates effective models where the spin-1 fermionic field is replaced by a composition of a Dirac fermion and the d-bein. Exploring a natural solution for topological actions, $\mathcal{F}=0$, we find non-trivial solutions for the fields. Specially, we find that the torsion plays an important role in terms of fermionic condensates. Analyzing the gauge transformations, we have shown that the Randall-Sundrum dimensional reduction respects the gauge transformation with the no-gravitini assumption. The 4-dimensional equations give us a unusual Ricci scalar dependence on the fermionic bilinears. A study of the fermionic behavior in the 4-dimensional brane is a next step in our endeavor, so that we may go deeper into the properties of our effective model we have discussed in the present contribution.
\paragraph*{}

\section*{Acknowledgments}
 This work was funded by the Brazilian National Council for Scientific and Technological Development (CNPq). 
\section*{Appendix}\label{app}
\paragraph*{}The representation of the generators is given in terms of $(4 + \mathcal{N}) \times (4 + \mathcal{N})$ supermatrices:\\
\\{
\begin{eqnarray}\nonumber
&&J_{ab}=\begin{bmatrix}
\frac{1}{2} (\gamma_{ab})^\alpha_{~\beta} & 0\\
0 & 0\\
\end{bmatrix} ~~,~~ J_{a}=\begin{bmatrix}
 (\gamma_{a})^\alpha_{~\beta} & 0\\
0 & 0\\
\end{bmatrix} ~~,~~ \\\nonumber
&& T_k=\begin{bmatrix}
0 & 0\\
0 & (\tau^k)_r^{~s}\\
\end{bmatrix} ~~,~~ 
Q^\alpha_s=\begin{bmatrix}
0 & 0\\
- \delta^r_s \delta^\alpha_\beta & 0\\
\end{bmatrix} ~~,~~ \\
&&\bar{Q}_\alpha^s=\begin{bmatrix}
0 & \delta^r_s \delta^\alpha_\beta\\
0 & 0\\
\end{bmatrix} ~~,~~ \mathbb{K}=\begin{bmatrix}
\frac{i}{4}\delta^\alpha_\beta & 0\\
0 & \frac{1}{\mathcal{N}}\delta^{~s}_r\\
\end{bmatrix}~.
\end{eqnarray}
}\\
\paragraph*{}From that, the following algebra can be written:
\begin{eqnarray}\nonumber
&&[J^{ab}, J^{cd}] = \eta^{ad}J^{bc} - \eta^{ac}J^{bd} + \eta^{bc}J^{ad} -\eta^{bd}J^{ac}  ~~ , ~~\\\nonumber
&& [J^a, J^b] = s^2 J^{ab} ~~,~~ [J^a, J^{bc}] =\eta^{ab} J^c - \eta^{ac} J^b~~,~~\\\nonumber
&&[J^a, Q_s]= -\frac{s}{2}\gamma^a Q_s ~~,~~ [J^a, \bar{Q}^s]=  \frac{s}{2}\bar{Q}^s\gamma^a ~~,~~\\\nonumber && [J^{ab}, Q_s] = -\frac{1}{2} \gamma^{ab} Q_s ~~,~~[J^{ab}, \bar{Q}^s] = \frac{1}{2} \bar{Q}^s\gamma^{ab} ~~,~~\\\nonumber&& [\mathbb{K}, Q_s] = -i(\frac{1}{4}- \frac{1}{\mathcal{N}}) Q_s ~~,~~ [\mathbb{K}, \bar{Q}^s] =  i(\frac{1}{4}- \frac{1}{\mathcal{N}}) \bar{Q}^s ~~,~~ \\\nonumber &&[T^k, Q_s] = (\tau^k)_s^r Q_r~~,~~ [T^k, \bar{Q}^s] = -(\tau^k)^s_r \bar{Q}^r\\\nonumber&& \{Q_s, \bar{Q}^r \} = - \frac{i}{2} \delta^r_s \gamma^a J_a - \frac{1}{4}\delta^r_s\gamma^{ab} J_{ab} + i \delta^r_s \mathbb{K} + (\tau^k)_s^r T_k ~.\\
\end{eqnarray}
\paragraph*{}All the other relations vanish. In the dimensionally reduced scenario, we have that the covariant derivative can be written as $\hat{\nabla} = (\nabla, \nabla_\chi d \chi)$, where:
\begin{eqnarray}\nonumber
\nabla_r^s &=&\Big{(} d +i(\frac{1}{4}- \frac{1}{\mathcal{N}})b + \frac{1}{2}  e_I \gamma^I + \frac{1}{2} e^4 \gamma_5 + \frac{1}{4} \omega_{IJ} \gamma^{IJ} +\\&+&  \frac{\lambda}{2}  b_I \gamma_5\gamma^{I}\Big{)}\delta_r^s +A_k (\tau^k)_r^{~s}~,
\end{eqnarray}
\begin{eqnarray}\nonumber
(\nabla_\chi)^s_r &=&\Big{(} \partial_\chi +i(\frac{1}{4}- \frac{1}{\mathcal{N}})b_\chi + \frac{1}{2}  (e_I)_\chi \gamma^I + \frac{1}{2} e^4_\chi \gamma_5 +\\\nonumber &+&  \frac{1}{4} (\omega_{IJ})_\chi \gamma^{IJ} +\frac{1}{2} \lambda (b_I)_\chi \gamma_5 \gamma^{I}\Big{)}\delta_r^s + (A_k)_\chi (\tau^k)_r^{~s}~,\\ 
\end{eqnarray}
\paragraph*{}In terms of the definitions of the dimensional reduction found in  \eqref{dimred1},\eqref{dimred2} and \eqref{dimred3}, we may rewrite the gauge transformations in terms of the component fields. For $\hat{e}^a$, we have:
\begin{subequations}
\begin{eqnarray}\nonumber
\delta e^I &=& d \epsilon^I + \omega^{IJ} \epsilon_J + \lambda b^I \epsilon_4 + \frac{1}{2}(\bar{\psi}^r (\Gamma + \gamma_5 e^4) \gamma^I \chi_r +\\&& +  \frac{1}{2}\bar{\chi}^r\gamma^I  (\Gamma + \gamma_5 e^4) \psi_r)~,
\end{eqnarray}
\begin{equation}
\delta e^4 = d \epsilon^4 + \lambda b^I \epsilon_I + \frac{1}{2}\bar{\psi}^r (\Gamma + \gamma_5 e^4) \gamma_5 \chi_r +  \frac{1}{2}\bar{\chi}^r\gamma_5  (\Gamma + \gamma_5 e^4) \psi_r~,
\end{equation}
\begin{eqnarray}\nonumber
\delta e^I_\chi&=& \partial_\chi \epsilon^I + \omega^{IJ}_\chi \epsilon_J + \lambda b^I_\chi \epsilon_4 + \frac{1}{2}\bar{\psi}^r (\gamma_J e^J_\chi + \gamma_5 e^4_\chi) \gamma^I \chi_r +  \\&& + \frac{1}{2}\bar{\chi}^r\gamma^I  (\gamma_J e^J_\chi + \gamma_5 e^4_\chi) \psi_r~,
\end{eqnarray}
\begin{eqnarray}\nonumber
\delta e^4_\chi &=& \partial_\chi \epsilon^4 +  \lambda b^I_\chi \epsilon_{I4} + \frac{1}{2}\bar{\psi}^r (\gamma_I e^I_\chi + \gamma_5 e^4_\chi) \gamma_5 \chi_r +  \\&& +\frac{1}{2}\bar{\chi}^r\gamma_5  (\gamma_I e^I_\chi + \gamma_5 e^4_\chi) \psi_r~,
\end{eqnarray}
\end{subequations}
\paragraph*{}where $\Gamma = e^I \gamma_I$. For $\hat{\omega}^{ab}$, it follows that:
\begin{subequations}
\begin{eqnarray}\nonumber
\delta \omega^{IJ} &=& d\epsilon^{IJ} + \omega^{[IK} \epsilon_K^{~~J]} +\frac{1}{4}\bar{\psi}^r (\Gamma + \gamma_5 e^4) \gamma^{IJ} \chi_r +  \\&& +\frac{1}{4}\bar{\chi}^r\gamma^{IJ} (\Gamma + \gamma_5 e^4) \psi_r  + \lambda b^{[I} \epsilon_4^{~J]}~,
\end{eqnarray}
\begin{eqnarray}\nonumber
\delta b^{I} &=&\frac{1}{\lambda} d\epsilon^{4I} + b^{K} \epsilon_K^{~~I} + \frac{1}{2\lambda}\bar{\psi}^r (\Gamma + \gamma_5 e^4) \gamma^{I}\gamma_5 \chi_r +\\&& +\frac{1}{2 \lambda} \bar{\chi}^r\gamma^{I} \gamma_5(\Gamma + \gamma_5 e^4) \psi_r~,
\end{eqnarray}
\begin{eqnarray}\nonumber
\delta \omega^{IJ}_\chi &=& \partial_\chi\epsilon^{IJ} + \omega_\chi^{[IK} \epsilon_K^{~~J]}  +\frac{1}{4}\bar{\psi}^r (\gamma_J e^J_\chi + \gamma_5 e^4_\chi) \gamma^{IJ} \chi_r + \\&& +\frac{1}{4} \bar{\chi}^r\gamma^{IJ} (\gamma_J e^J_\chi + \gamma_5 e^4_\chi) \psi_r + \lambda b_\chi^{[I} \epsilon_4^{~J]}~,
\end{eqnarray}
\begin{eqnarray}\nonumber
\delta b^I_\chi &=&\frac{1}{\lambda} \partial_\chi \epsilon^{4I} + b_\chi^K \epsilon_{K}^{~~I} + \frac{1}{2\lambda}\bar{\psi}^r (\gamma_J e^J_\chi + \gamma_5 e^4_\chi) \gamma^{I} \gamma_5\chi_r + \\&& + \frac{1}{2 \lambda}\bar{\chi}^r\gamma^{I} \gamma_5(\gamma_J e^J_\chi + \gamma_5 e^4_\chi) \psi_r~,
\end{eqnarray}
\end{subequations}

\paragraph*{}For $\psi_r$, one gets:
\begin{subequations}
\begin{equation}
\delta \Big{[}(\Gamma+ \gamma_5 e^4)\psi_r \Big{]}= \vec{\nabla}\chi_r~,
\end{equation}

\begin{equation}
\delta \Big{[} (\gamma_J e^J_\chi + \gamma_5 e^4_\chi) \psi_r \Big{]} = \vec{\nabla}_\chi \chi_r~.
\end{equation}
\end{subequations}

\paragraph*{}And, finally, for $b$, it follows that:
\begin{subequations}

\begin{equation}
\delta b = d \epsilon_b + i \bar{\psi}^r (\Gamma + \gamma_5 e^4)  \chi_r +  i\bar{\chi}^r (\Gamma + \gamma_5 e^4) \psi_r~,
\end{equation}
\begin{eqnarray}\nonumber
\delta b_\chi &=& \partial_\chi \epsilon_b + i\bar{\psi}^r (\gamma_J e^J_\chi + \gamma_5 e^4_\chi)  \chi_r + \\&& + i\bar{\chi}^r (\gamma_J e^J_\chi + \gamma_5 e^4_\chi) \psi_r~.
\end{eqnarray}
\end{subequations}

\paragraph*{}Going on with our analysis, we write down the field-strength in terms of components. First, for the $F^a$ components, we have:
\begin{subequations}
\begin{equation}
F^I = T^I + \lambda b^I e_4 + \bar{\psi}^r (\Gamma + \gamma_5 e^4)\gamma^I (\Gamma + \gamma_5 e^4) \psi_r~,
\end{equation}
\begin{eqnarray}\nonumber
F^I_\chi &=& \partial_\chi e^I + \omega_\chi^{IJ} e_J - D(\omega)e_\chi^I + \\&&+ \bar{\psi}^r (\Gamma + \gamma_5 e^4)\gamma^I (\gamma_J e_\chi^J + \gamma_5 e^4_\chi) \psi_r~,
\end{eqnarray}
\begin{equation}
F^4 = d e^4 + \lambda b^I e_I + \bar{\psi}^r (\Gamma + \gamma_5 e^4)\gamma_5 (\Gamma + \gamma_5 e^4) \psi_r~,
\end{equation}
\begin{eqnarray}\nonumber
F^4_\chi &=& \partial_\chi e^4 + \lambda b_\chi^I e_I - de_\chi^4 - \lambda b_I e_\chi^I + \\&&+\bar{\psi}^r (\Gamma + \gamma_5 e^4)\gamma_5 (\gamma_J e_\chi^J + \gamma_5 e^4_\chi) \psi_r~.
\end{eqnarray}

\end{subequations}
\paragraph*{}
\paragraph*{}Finally, for $\hat{F}^{ab}$, we find:
\begin{subequations}
\begin{eqnarray}\nonumber
F^{IJ} &=& R^{IJ} + e^I e^J + \lambda^2 b^I b^J + 
\\&&+\frac{1}{2}\bar{\psi}^r (\Gamma + \gamma_5 e^4)\gamma^{IJ} (\Gamma + \gamma_5 e^4) \psi_r~,
\end{eqnarray}
\begin{eqnarray}\nonumber
F_\chi^{IJ} &=& d\omega_\chi^{IJ}+ \omega^{I}_{~L} \omega_\chi^{LJ} - \partial_\chi \omega^{IJ} + \lambda^2 b_\chi^I b^J + e_\chi^I e^J + \\&&+\frac{1}{2}\bar{\psi}^r (\Gamma + \gamma_5 e^4)\gamma^{IJ} (\gamma_K e^K_\chi + \gamma_5 e_\chi^4) \psi_r ~,
\end{eqnarray}
\begin{eqnarray}\nonumber
F^{I4} &=& \lambda D(\omega)b^I + e^I e^4 + \bar{\psi}^r (\Gamma 
+ \gamma_5 e^4)\gamma^{I} \gamma_5(\Gamma + \gamma_5 e^4) \psi_r~,\\
\end{eqnarray}
\begin{eqnarray}\nonumber
F^{I4}_\chi &=& \lambda(D(\omega) b_\chi^I - \partial_\chi b^I - \omega_\chi^{IJ} b_J) + e_\chi^I e^4 - e^I e_\chi^4 + \\&&+\bar{\psi}^r (\Gamma + \gamma_5 e^4)\gamma^{I} \gamma_5(\gamma_J e_\chi^J + \gamma_5 e^4_\chi) \psi_r~.
\end{eqnarray}
\end{subequations}
\paragraph*{}For the gauge fields, we find:
\begin{subequations}

\begin{eqnarray}\nonumber
F^k &=& d A^k +  f^{k}_{~l m}A^{l} A^m  +\\&&+ \bar{\psi}^r(\Gamma + \gamma_5 e^4) (\tau^k)_r^{~s}(\Gamma + \gamma_5 e^4)\psi_s~,
\end{eqnarray}
\begin{eqnarray}\nonumber
F_\chi^k &=& \partial_\chi A^k +  f^{k}_{~lm}A_\chi^{l} A^m - d A^k_\chi+\\\nonumber &&+ \bar{\psi}^r(\Gamma + \gamma_5 e^4)(\tau^k)_r^{~s}(\gamma_J e_\chi^J + \gamma_5 e^4_\chi) \psi_s~,\\
\end{eqnarray}
\begin{equation}
F =  d b + i \bar{\psi}^r(\Gamma + \gamma_5 e^4)(\Gamma + \gamma_5 e^4) \psi_r~,
\end{equation}
\begin{equation}
F_\chi =  d b_\chi -\partial_\chi b + i \bar{\psi}^r(\Gamma + \gamma_5 e^4)(\gamma_J e_\chi^J + \gamma_5 e^4_\chi)\psi_r
\end{equation}

\end{subequations}
\paragraph*{}By adapting the RS ansatz, we find the solutions which we have presented in Section \ref{RSR}.

\end{document}